\documentclass{svmult}
\usepackage{dsfont}
\usepackage{fancyhdr}
\usepackage{color}
\usepackage{eurosym}
\usepackage[dvips]{graphicx}
\usepackage{pslatex}

\begin{document}
\title*{Levy-Lieb Principle meets Quantum Monte Carlo}
\author{Luigi Delle Site}
\institute{Luigi Delle Site \at Institute for Mathematics, Freie Universit\"{a}t
Berlin, Arnimallee 6, D-14195 Berlin, Germany\\
\email{luigi.dellesite@fu-berlin.de}}
\maketitle
\abstract{ We review an approach where the energy functional of Density-Functional
Theory (DFT) can be determined without empiricism via a Quantum Monte Carlo (QMC)
procedure. The idea consists of a nested iterative loop where the configurational
space of the electrons in the QMC procedure is determined by a trial one-particle
electron density; this allows then for the determination of a first guess for the
energy functional, which is then in turn used for a DFT minimization process. This
latter delivers a new electron density and the loop is repeated iteratively until
convergence is reached and thus the electronic structure of the system determined.
In essence, from the applied point of view,  
such an approach is a compromise between the large computational cost but high
accuracy of QMC and the reasonable computational cost but lower accuracy of DFT. The
proposed approach goes beyond the standard Kohn-Sham method, which is the standard
in DFT, and requires neither orbitals nor
 the a priori specification of the exchange and correlation functional. From the
conceptual point of view, the relevant aspect is that the 3N-dimensional
wavefunction is derived as a function(al) of the three-dimensional electron
density, which is the key principle of DFT.}
\section{Introduction: The Need for going beyond standard Density Functional Theory}
The Hohenberg-Kohn (HK) formulation of Density Functional Theory (DFT)
\cite{h-k} proves the existence of a variational problem for the ground state of a
many-electron system: $E_{GS}=\min_{\rho}E[\rho]$; where $E_{GS}$ is the ground
state energy, $E[\rho]$ the electronic energy functional of the one particle density
$\rho({\bf r})$. It follows that a prohibitive multidimensional problem is reduced
to the search of a computationally accessible three-dimensional quantity, $\rho({\bf
r})$; this is the simple though revolutionary essence of DFT.
The energy functional $E[\rho]$ differs from system to system according to the
external potential but contains a universal term common to all electronic systems.
The energy functional is usually written as: $E[\rho]=F[\rho]+\int v({\bf
r})\rho({\bf r})$, with $v({\bf r})$ being the external potential
(e.g. nuclei-electron Coulomb interaction) and the universal
functional: $F[\rho]=T[\rho]+V_{ee}[\rho]$. $F[\rho]$, whose existence and
uniqueness is proved by the HK theorems, is composed by two
terms, the kinetic functional $T[\rho]$ and the electron-electron Coulomb
functional $V_{ee}[\rho]$. Unfortunately the form of $T[\rho]$ and
$V_{ee}[\rho]$ is unknown. One solution to this
problem was later on provided by Kohn and Sham and nowadays is known as the
Kohn-Sham (KS) approach \cite{k-s,kmphys}. The essential features of the KS approach
are the following : it introduces $\frac{N}{2}$ single particle
orbitals in a non interacting frame, $\phi_{i}({\bf r})$, each accommodating two
electrons according to
the Pauli principle. It follows that: $\rho({\bf
  r})=\sum_{i=1}^{\frac{N}{2}}|\phi_{i}({\bf r})|^{2}$; this in turn leads to the
simplification of the kinetic functional $T[\rho]$ which now can be written
exactly in its non interacting form
$T_{s}[\phi]=\sum_{i=1}^{\frac{N}{2}}|\nabla |\phi_{i}({\bf r})|^{2}$. 
Also $V_{ee}[\rho]$ is simplified and reduced to the (classical) Hartree term,
$V_{Hartree}[\rho]=\int\int\frac{\rho({\bf r})\rho({\bf r^{'}})}{|{\bf r}-{\bf
r^{'}}|}d{\bf r}d{\bf r^{'}}$.
Next the unknown part of $F[\rho]$, due to the missing interacting part of the
orbitals, is contained in
the so called exchange and correlation energy term,
$E_{xc}[\rho]=T_{corr}[\rho]+V_{xc}[\rho]$; with $T_{corr}[\rho]$ the correlation
term of the kinetic functional and $V_{xc}[\rho]$ the exchange and correlation term
coming from the electron-electron Coulomb interaction. We obtain then:
$E[\rho]=T_{s}[\rho]+T_{corr}[\rho]+V_{Hartree}[\rho]+V_{xc}[\rho]$.
This setting reduces the HK variational problem to a system of $\frac{N}{2}$
Schr\"{o}dinger-like equations: $\frac{\hbar^{2}}{2m}\nabla^{2}\phi_{i}({\bf
  r})+v_{eff}(\rho,{\bf r})\phi_{i}({\bf r})=\epsilon_{i}\phi_{i}({\bf r})$.
Here, $\hbar$ is Planck's constant, $m$ the electron mass, $\epsilon_{i}$
the equivalent of an eigenvalue for the $i$-th orbital. Finally, the effective
potential, 
$v_{eff}(\rho,{\bf r})= v({\bf r})+\int\frac{\rho({\bf r^{'}})}{|{\bf r}-{\bf
  r^{'}}|}d{\bf r^{'}}+v_{xc}(\rho,{\bf r})$, with $v_{xc}(\rho,{\bf
  r})=\frac{\delta E_{xc}[\rho]}{\delta\rho}$ (see e.g. \cite{yparr}).\\
Differently from $F[\rho]$, $E_{xc}[\rho]$ can be derived under specific conditions
as it is the case of local density and generalized gradient approximation (LDA and
GGA) \cite{primer}. The chapter of Ghiringhelli reports the success of the
KS-LDA/GGA for real systems but enumerates also its large number of failures. Modern
technology based on atomic control of physical processes, requires an accuracy which
is much beyond that provided by standard LDA/GGA approaches of DFT; for example,
standard GGA cannot properly describe van der Waals interactions, as underlined in
the chapter of Watermann {\it et al.}. This is a key aspect for understanding and predicting
the behaviour of many systems, in particular those of biological nature
\cite{dnamatthias}. The urging demand to develop approaches to include van der Waals
in DFT, has generated during the years a  whole industry devoted to building
approximate methods for practical (numerical) and/or system specific, though often
rather expensive, solutions \cite{alex, chalmers, dyson}. However, so far, a
universal strategy transferable to any system and any physical situation has not
been found \cite{savin}. In general, the standard paths followed by the community in
search of accurate $E_{xc}[\rho]$, is that of climbing Perdew's ``Jacobs Ladder''
\cite{primer,banesko} (for example following the approach of the Random Phase
Approximation \cite{rinke}) or by applying straightforward numerical
parameterizations \cite{truhlar}. These strategies however add a level of complexity
to the theory which brings DFT far away from its character of simple and yet
rigorous approach; that is the characteristic that made DFT popular. In practice the
wished ``{\it leap forward}'', expected over the last twenty years,  did not really
happen yet. For practical purposes  and  for short term computational strategies,
the current standard paths to $E_{xc}[\rho]$ are indisputably very useful, however
for long term strategies, with the demand of increasing accuracy to be faced in the
next decades, a major effort should be done to develop conceptually more satisfying,
efficient and universal theoretical/numerical frameworks. These latter must not
necessarily be substitutive of the current strategies but actually must complement
them; such a complementarity would optimize the balance between immediate
necessities of practical applications and long terms development of conceptually
more satisfying functionals. In this context one possible long term strategy would
be that of going back to the fundamentals of DFT, use as much as possible the known
mathematical properties of the functional and combine them with the essential
principles of quantum mechanics without having the homogeneous non-interacting
electron gas as a starting paradigm; the latter should rather be a mere (necessary)
validating limiting case. Mathematical research along this direction has made many
progresses, though often not fully shared by those working on more practical and
applicative aspects of DFT (see e.g. chapters of Siedentop and Bach). In fact, for
practical applications, the most delicate step of this strategy is related to the
optimal inclusion of formal/conceptual results in efficient computational protocols.
If the balance between computational costs and conceptual rigor (that is
physical/numerical accuracy) would be positive (according to the standards of the
field) than this sort of strategy would probably represent the  path to the
auspicated ``{\it leap forward}''. The current chapter represents in this context
the description of an attempt to explore a path along the idea of including as much
as possible mathematical analytical results and combine them with basic physical
principles of quantum mechanics for a feasible computational protocol. The essential
idea is based on a reformulation of the Levy-Lieb (LL) principles of DFT
\cite{ll1,ll2} in terms of electron density and $N-1$-conditional probability
density. This latter can then be formally treated within the computational framework
of the Ground State Path Integral Quantum Monte Carlo (GSPI QMC). The resulting
scheme represents a conceptually well founded hybrid framework merging DFT and QMC
so that the energy functional is determined on-the-fly during a DFT calculation. In
more practical terms, this essentially represents a compromise between the accuracy
but high costs of QMC (see chapter of Sorella) and the feasibility but
lower accuracy of standard DFT. Interestingly, since it can be applied to any
system, one may think of it as an implicit numerical definition of the universal
functional $F[\rho]$.

\section{Starting point: Reformulation of the Levi-Lieb Constrained-Search Principle}
Let us start by specifying one simple but crucial formal definition; following the
idea of Sears, Parr and Dinur \cite{sears}, we decompose the $3N$-dimensional
probability density of a system of $N$ interacting electrons (i.e. the square of its
ground state wavefunction) as: $|\psi({\bf r}_{1},{\bf r}_{2},....{\bf
r}_{N})|^{2}=\rho({\bf r}_{1})f({\bf r}_{2},...{\bf r}_{N}|{\bf r}_{1})$. Here ${\bf
r}_{i}$ indicates the space coordinate of the $i$-th electron, and spins are not
explicitly considered, $\rho({\bf r}_{1})=\int|\psi({\bf r}_{1}, {\bf r}_{2},...{\bf
r}_{N})|^{2}d{\bf r}_{2}....d{\bf r}_{N}$ is the one particle electron density and
is such that $\int\rho({\bf r}_{1})d{\bf r}_{1}=N$, finally $f({\bf
r}_{2},..........{\bf r}_{N}|{\bf r}_{1})$ is the $N-1$-conditional probability
density.
The original formulation of the LL constrained search formulation (that is the very
rigorous generalization and validation of the Hohenberg-Kohn theorem of DFT)
\cite{ll1,ll2,yparr} is written in terms of wavefunctions:
\begin{equation}
E_{GS}=\min_{\rho}\left[\min_{\psi\to\rho}\langle\psi|K+V_{ee}|\psi\rangle+\int\rho({\bf
r})v({\bf r})d{\bf r}\right]
\label{ll-or}
\end{equation}
where $K$ and $V_{ee}$ are the kinetic and electron-electron Coulomb operator
respectively.
The inner minimization is done on the whole space of antisymmetric wavefunctions
that integrate to $\rho$ while the outer minimization searches for $\rho$ which
minimizes the global functional. The universal functional of Hohenberg and Kohn is
then rigorously defined as:
\begin{equation}
F[\rho]=\min_{\psi\to\rho}\langle\psi|K+V_{ee}|\psi\rangle
\label{univ-or}
\end{equation}  
These are all the essential ingredients of DFT. Starting, from here we can then
develop numerical procedures with different level of accuracy and numerical
efficiency.
For this chapter, the starting point is a reformulation of the LL principle in terms
of $f({\bf r}_{2},...{\bf r}_{N}|{\bf r})$ instead of $\psi({\bf r},{\bf
r}_{2},...{\bf r}_{N})$.
 Such a reformulation leads to a minimization problem on $f$ (see e.g.
\cite{lui2,lui3,lui4,lui5}): 
\begin{eqnarray}
\min_{f}\left(\Gamma[\rho,f]\right)=\min_{f}\left(\frac{1}{8}\int \rho({\bf
  r}_{1}) I({\bf r}_{1})d{\bf r}_{1}+(N-1)\int \rho({\bf
  r}_{1})C({\bf r}_{1})d{\bf r}_{1}\right)
\label{gamma}
\end{eqnarray}
where 
\begin{equation}
I({\bf r}_{1})=\int_{{\bf R}^{N-1}}\frac{|\nabla_{{\bf r}_{1}}f({\bf r}_{2},....{\bf
  r}_{N}|{\bf r}_{1})|^{2}}{f({\bf r}_{2},....{\bf
  r}_{N}|{\bf r}_{1})}d{\bf r}_{2}....d{\bf r}_{N}
\label{defi}
\end{equation}
and
\begin{equation}
C({\bf r}_{1})=\int_{{\bf R}^{N-1}}\frac{f({\bf r}_{2},....{\bf
  r}_{N}|{\bf r}_{1})}{|{\bf r}_{1}-{\bf r}_{2}|}d{\bf r}_{2}....d{\bf r}_{N}.
\label{defc}
\end{equation}
It must be noticed that antisymmetry in $f$ is taken into account in an indirect way, that is when two particles have same position (and same spin, here not explicitly treated), then $f$ is equal zero (see \cite{lui2,lui3,lui4,lui5}).
 The solution of the minimization in (\ref{gamma}) automatically leads to a
complete and exact energy functional and, as a consequence, to a full minimization
problem for the $\rho$ of the ground state:
\begin{equation}
E_{0}=\min_{\rho}\left[\left(\min_{f}\Gamma[\rho,f]\right)+\frac{1}{8}\int\frac{|\nabla\rho({\bf
      r}_{1})|^{2}}{\rho({\bf r}_{1})}d{\bf r}_{1}+\int v({\bf r}_{1}) d{\bf
    r}_{1}\right].
\label{eqg1}
\end{equation}
In the next section we will identify ${\bf r}_{1}$ with ${\bf r}$, so that we make
our formalism compatible with that used in standard derivations of DFT; moreover, it
must be noticed that due to the particles' indistinguishability,
the expression ${|{\bf r}_{1}-{\bf r}_{2}|}$ in (\ref{defc}) is equivalent to the
more familiar and general expression  ${|{\bf r}-{\bf r}^{'}|}$.   

\subsection*{Energy Density of the Ground State}
If we search for the ground state of a specific system of $N$ 
electrons with a well-defined external potential, then the procedure of inner
minimization of (\ref{eqg1}) (i.e. the  search for $f$ which minimizes
$\Gamma[\rho,f]$ w.r.t. $\rho$) leads to $f_{\min}=f_{GS}$.
Since $f_{\min}=f_{GS}$, the explicit expression of the functional $F[\rho]$ in the
ground state can be written as:
\begin{equation}
F[\rho_{GS}]=\int\rho_{GS}({\bf r})\left[\frac{1}{8}\frac{|\nabla\rho_{GS}({\bf
r})|^{2}}{\rho_{GS}({\bf r})^{2}}+\frac{1}{8}I_{f_{GS}}({\bf
r})+(N-1)C_{f_{GS}}({\bf r})\right]d{\bf r}.
\label{deff}
\end{equation}
The term:
\begin{equation}
\epsilon({\bf r})=\frac{1}{8}\frac{|\nabla\rho_{GS}({\bf r})|^{2}}{\rho_{GS}({\bf
r})^{2}}+\frac{1}{8}I_{f_{GS}}({\bf r})+(N-1)\, C_{f_{GS}}({\bf r})
\label{deff1}
\end{equation}
is an energy density per particle expressed in terms of its kinetic
$\left(\frac{1}{8}\frac{|\nabla\rho_{GS}({\bf r})|^{2}}{\rho_{GS}({\bf
r})^{2}}+\frac{1}{8}I_{f_{GS}}({\bf r})\right)$ and Coulomb $\left((N-1)\,
C_{f_{GS}}({\bf r})\right)$ parts.\\
Here, $I_{f_{GS}}({\bf r})$ and $C_{f_{GS}}({\bf r})$ indicate that the
quantities of (\ref{defi}) and (\ref{defc}) are calculated for the $f$
of the ground state.\\
If one knew $I_{f_{GS}}({\bf r})$ and $C_{f_{GS}}({\bf r})$ in terms of
$\rho_{GS}$, this would correspond to have the universal functional of
Hohenberg and Kohn. At this point the key question is whether there is any rigorous
technique to calculate, in practical terms (i.e., not only formally),
$f_{GS}$ and thus determine $I_{f_{GS}}({\bf r})$ and
$C_{f_{GS}}({\bf r})$.
 
\section{A Rigorous Scheme for $f$ via Ground State Path Integral Quantum Monte Carlo}
In the next sections we will describe a theoretical/computational protocol
\cite{ijqc} based on the use of the Ground State (GS) Path Integral (PI) Quantum
Monte
Carlo (QMC) \cite{pigs1,pigs2,pigs3} for the calculation of the energy densities via
$f_{GS}$, as derived in the previous section. Such a computational protocol
considers $f$ an implicit functional of the electron density, independently of the
external potential, thus, in principle, the ideal combination of the conceptual
framework of DFT and the numerical, feasible, calculation of QMC.

\subsection*{Ground State Path Integral Quantum Monte Carlo Method}
Within the framework of GSPI QMC the quantum partition function
of a system of $N$ particles can be written as:
\begin{equation}
Z=\int\psi({\bf R}_{0})\exp[-S({\bf R}_{0}, {\bf
    R}_{1}....{\bf R}_{M})]\psi({\bf R}_{M})d{\bf R}_{0}.....d{\bf R}_{M}
\label{partfu}
\end{equation}
${\bf R}_{0}=({\bf r}^{0},{\bf r}^{0}_{2},.....{\bf r}^{0}_{N}) \in \mathds{R}^{3N}$ is a
configuration of the $N$ particles in space while ${\bf R}_{1}\in \mathds{R}^{3N}$ is
another configuration and any other ${\bf R}_{m}$ is a different configuration. The
sequence ${\bf
  R}_{0}.....{\bf R}_{M}$ is an open path of length $M$ in the spaces of the
$N$-particle configurations. $\psi({\bf R}_{0})$ and $\psi({\bf R}_{M})$ are the values of 
a trial wavefunction calculated at the initial and final configuration.  
The action $S({\bf R}_{0}, {\bf R}_{1}....{\bf R}_{M})$ is defined as:
\begin{equation}
\exp[-S({\bf R}_{0}, {\bf R}_{1}....{\bf R}_{M})]=\left\langle{\bf R}_{0}|e^{-\tau H}|{\bf
R}_{1}\right\rangle\left\langle{\bf R}_{1}|e^{-\tau H}|{\bf R}_{2}\right\rangle.....
\left\langle{\bf R}_{M-1}|e^{-\tau H}|{\bf R}_{M}\right\rangle
\end{equation}
where $\tau$, formally an imaginary timestep, is defined as: $\tau=\frac{\beta}{M}$,
where $\beta$, formally a Boltzmann factor is practically a parameter that regulates
the convergence, and $H$ is the Hamiltonian. 
The quantum mechanical partition function becomes an integral involving a sequence
of transitional probabilities in imaginary time $t=\beta/2$. In turn, each of these
transition probabilities can be factorized into a kinetic part:
\begin{equation}
\left\langle{\bf R}_{i}|e^{-\tau K}|{\bf
R}_{i+1}\right\rangle=\frac{1}{(2\pi\tau)^{3N/2}}e^{-\frac{\tau}{2}\left(\frac{{\bf
R}_{i}-{\bf R}_{i+1}}{\tau}\right)^{2}}
\label{spring}
\end{equation}
and a potential part:
\begin{equation}
\left\langle{\bf R}_{i}|e^{-\tau V}|{\bf
R}_{i+1}\right\rangle=\frac{1}{(2\pi\tau)^{3N/2}}e^{-\frac{\tau}{2}[V({\bf R}_{i})+V({\bf
R}_{i+1})]}
\label{pot}
\end{equation}
with $V$ being the potential operator of the system considered.
For atoms and molecules $V({\bf R})=V_{ee}+V_{ne}$, that is the electron-electron
and the nucleus-electron interaction.
For electrons (fermions), for a real $\psi({\bf R})$, the fixed node
condition is usually employed (see \cite{pigs1}):
\begin{eqnarray}
V_{fermions}({\bf R})=V({\bf R})~~for~~~\psi({\bf R})>0 \\
V_{fermions}({\bf R})=\infty~~~~~~~for~~~\psi({\bf R})\le 0.
\end{eqnarray}
For complex $\psi({\bf R})$, a term is added to the free-particle part
of the action. 
Being the wavefunction defined as: $\psi_{t}({\bf R})=e^{-t H}\Psi({\bf R})$,
with $\Psi({\bf R})$ the ground state wavefunction, for $t\to\infty$ (i.e. large number of time steps $\tau$, or equivalently, large number of $M$), $\psi\to\Psi({\bf R})$. 
In practice, the wavefunction is calculated at the midpoint of the path, i.e. at
${\bf R}_{M/2}$.\\
From the computational point of view the calculation of the quantities involved in
the problem is reduced to (mapped onto) the sampling of configurations of a system
of $N$ linear polymers of length $M$ at a fictitious temperature $\tau$, with
polymer-polymer interactions restricted to interactions between beads of different
polymers but with the same label.

\subsection*{Determination of $f$ and $\Gamma$ via GSPI QMC}
In order to proceed in the derivation of $f_{GS}$ and $\Gamma_{GS}$ of (\ref{gamma}) in terms
of the GSPI approach, we adopt the following convention:
The configuration at the midpoint of the path, ${\bf R}_{M/2}$ , is indicated as
${\bf R}_{*}$. This means that $({\bf r}^{M/2},{\bf r}_{2}^{M/2},........{\bf
r}_{N}^{M/2})$ becomes $({\bf r}^{*},{\bf r}^{*}_{2},........{\bf r}^{*}_{N})$.
 According to (\ref{partfu}), the $N-1$-conditional probability density $f_{GS}$
 can now be written as:
\begin{eqnarray}
\label{effe1}
f({\bf r}^{*}_{2},....{\bf r}^{*}_{N}|{\bf r}^{*})=\frac{1}{Z_{{\bf r}^{*}}}\int
d{\bf R}_{0}d{\bf R}_{1}.......d{\bf R}_{\frac{M}{2}-1}d{\bf
R}_{\frac{M}{2}+1}.......d{\bf R}_{M} 
\\ \psi({\bf R}_{0})\exp[-S({\bf
R}_{*}, {\bf
    R}_{0},{\bf R}_{1},....{\bf R}_{\frac{M}{2}-1},{\bf R}_{\frac{M}{2}+1}......{\bf
R}_{M})]\psi({\bf R}_{M})\nonumber
%\label{effe1}
\end{eqnarray}
where:
\begin{eqnarray}
Z_{{\bf r}^{*}}=\int \exp[-S({\bf R}_{*},{\bf
R}_{0},...{\bf R}_{\frac{M}{2}-1},{\bf R}_{\frac{M}{2}+1},..{\bf R}_{M})]\psi({\bf
R}_{M})\\d{\bf R}_{*}^{N-1}d{\bf R}_{0}....d{\bf R}_{\frac{M}{2}-1}d{\bf
R}_{\frac{M}{2}+1}...d{\bf R}_{M} \psi({\bf R}_{0})\nonumber
\label{zetaf}
\end{eqnarray}
$d{\bf R}_{*}^{N-1}$ means that the integration is done on the whole space of
configurations ${\bf r}^{*}_{2},...{\bf r}^{*}_{N}$ of ${\bf R}_{*}$ except
that corresponding to variable ${\bf r}^{*}$.
With this set up, $f$ can be calculated by propagating
stochastically, according to a Monte Carlo procedure, the path ${\bf R}$ in
imaginary time with timestep $\tau$.
Since the GSPI procedure, when evaluating in ${\bf R}_{*}$, delivers the ground
state wavefunction of the system,
the expression of  $f$ in (\ref{effe1}) corresponds to the ground state
$N-1$-conditional probability density, 
that is, it corresponds to $f_{\min}$  (or better to $\min_{f}$) of (\ref{gamma}).
The expression of (\ref{effe1}) then leads to:
\begin{equation}
I({\bf r}^{*})=\int_{{\bf R}^{N-1}}\frac{|\nabla_{{\bf r}^{*}}f({\bf
r}^{*}_{2},....{\bf
  r}^{*}_{N}|{\bf r}^{*})|^{2}}{f({\bf r}^{*}_{2},....{\bf
  r}^{*}_{N}|{\bf r}^{*})}d{\bf r}^{*}_{2}....d{\bf r}^{*}_{N}
\label{irho}
\end{equation}
and
\begin{equation}
C({\bf r}^{*})=\int_{{\bf R}_{N-1}}\frac{f({\bf r}^{*}_{2},....{\bf
  r}^{*}_{N}|{\bf r}^{*})}{|{\bf r}^{*}-{\bf r}^{*}_{2}|}d{\bf r}^{*}_{2}....d{\bf
r}^{*}_{N}.
\label{crho}
\end{equation}
Where now $I_{f_{\min}}({\bf r})=I({\bf r}^{*})$ and $C_{f_{\min}}({\bf r})=C({\bf
r}^{*})$.
The Hohenberg-Kohn functional in {\it local} form becomes:
\begin{equation}
F[\rho]=\frac{1}{8}\int\frac{|\nabla\rho({\bf
      r}^{*})|^{2}}{\rho({\bf r}^{*})}d{\bf r}^{*}+\frac{1}{8}\int \rho({\bf
r}^{*})I({\bf r}^{*})d{\bf r}^{*}+(N-1)\int\rho({\bf r}^{*})C({\bf
r}^{*})d{\bf r}^{*}.
\label{fhk}
\end{equation}

\subsection*{On-the-fly derivation of the energy functional}
\begin{figure}[h!]
\centering
\includegraphics[width=1.0\textwidth]{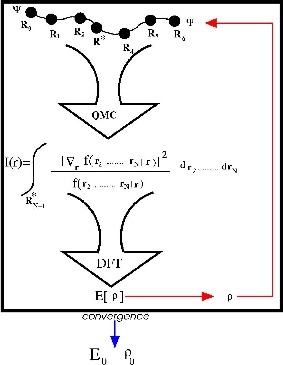}
\caption{Pictorial scheme of the DFT-QMC. The QMC step allows for the calculation of
a energy density at a given $\rho_{i}$; such an energy density is taken as an
approximate guess for the energy functional of DFT. Next, $\rho_{i}$ is updated via
a DFT minimization where together with the guessed energy functional the external
potential is explicitly added. The minimization delivers a density $\rho_{i+1}$ and
this latter is taken as a pre-selection criterion in generating a random walk in the
configuration space of the QMC. This updates the energy functional which now can be
used for a further DFT minimization which will deliver a density $\rho_{i+2}$; the
procedure continues then iteratively until convergence is reached.} 
\label{fig}
\end{figure}

The formal apparatus derived in the previous sections allows now to design a
combined DFT-QMC iterative procedure by which the energy functional is derived {\it
on-the-fly} during a standard DFT minimization.
The key point consists in modifying the GSPI
approach so that the resulting $f$ is a functional of $\rho$ only,
independently from the external potential $v({\bf r})$.
To this end we rewrite the transitional probability for the potential part as:
\begin{equation}
\left\langle{\bf R}_{i}|e^{-\tau V}|{\bf
R}_{i+1}\right\rangle=\frac{1}{(2\pi\tau)^{3N/2}}e^{-\frac{\tau}{2}[V_{ee}({\bf
R}_{i})+V_{ee}({\bf R}_{i+1})]}
\label{pot3}
\end{equation}
i.e., considering only the electron-electron interaction. The transitional
probability for the kinetic part remains the same of ( \ref{spring}).
Next, we can calculate $f$, and thus $I({\bf r})$ and $C({\bf r})$ as in
(\ref{effe1}),
(\ref{irho}) and (\ref{crho}), but with a sampling restricted to a trial density
$\rho_{trial}$. This means sampling the ${\bf R}_{i}'s$ in the configuration space
with the constraint that the one-particle density is $\rho_{trial}$. 
In this case the QMC procedure assures that the principle:
$\min_{f}\Gamma[\rho,f]$, is achieved in the sense that the resulting $\Gamma$
is that of {\it  ``ground state''} (i.e. having minimal kinetic plus pair potential energy) at  fixed $\rho_{trial}$.
Note that at this stage the external potential is not invoked and it is actually
absent; in practice what we have done is to find $f$ of ground state of a gas of
electrons with some {\it artificially forced} electron density. 
From the obtained $f$ we can now calculate the corresponding $I({\bf
r})=I_{\rho_{trial}}({\bf r})$ and $C({\bf
  r})=C_{\rho_{trial}}({\bf r})$. These quantities are taken as a first guess for an
analytic (or numerical) fitting in order to write an energy functional for a
generic $\rho({\bf r})$:
\begin{equation}
E[\rho]=\int\rho({\bf r})\left[\frac{1}{8}\frac{|\nabla\rho({\bf r})|^{2}}{\rho({\bf
r})^{2}}d{\bf
  r}+\frac{1}{8}I_{\rho_{trial}}({\bf r})+(N-1)C_{\rho_{trial}}({\bf r})+v({\bf
    r})\right]d{\bf r}.
\label{funtrial}
\end{equation}
Next, we use (\ref{funtrial}) for a minimization
w.r.t. $\rho$ and obtain a new $\rho=\rho^{1}_{out}$, different from
$\rho_{trial}$ because in (\ref{funtrial}) the effect of $v({\bf r})$ is explicitly
included during the energy minimization.
Note that $\rho^{1}_{out}$ is the equivalent of $\rho_{GS}$ corresponding to the approximate functional of (\ref{funtrial}).\\
At this point one can use $\rho^{1}_{out}$ as a new trial density, repeat the 
QMC procedure, that is we search the ground state of a gas of electrons with 
{\it artificially forced} electron density $\rho^{1}_{out}$, this will lead to a new
$f$ and in turn to a new $I({\bf r})$ and $C({\bf r})$ and thus we can have a new
guess for $E[\rho]:\int\rho({\bf r})\left[\frac{1}{8}\frac{|\nabla\rho({\bf
r})|^{2}}{\rho({\bf r})^{2}}d{\bf
  r}+\frac{1}{8}I_{\rho^{1}_{out}}({\bf r})+(N-1)C_{\rho^{1}_{out}}({\bf r})+v({\bf
    r})\right]d{\bf r}.$ As above we can then use the expression $E[\rho]$ (again
for a generic $\rho({\bf r})$) for a minimization w.r.t. $\rho({\bf r})$ and
obtain as a result a new $\rho^{2}_{out}$ and repeat the procedure until
$\rho^{i}_{out}=\rho^{i+1}_{out}$ with some accuracy. Figure \ref{fig} provides
a pictorial representation of the scheme. Essentially, the procedure can be seen
from two different points of view:
\begin{itemize}
\item From the QMC: the {\it a priori} knowledge of $\rho$ can be used as a fast
pre-selection criterion in choosing the space configurations within the MC sampling.
The effect of the external potential is taken care by the minimization process. 
\item From the DFT: There is no need for any system to specify the functional {\it a
priori}, this is calculated in a very accurate way on-the-fly during the
minimization process. Moreover, any calculation will provide new data to be used for
testing or improving existing functionals.
\end{itemize}

It must be noticed an important conceptual aspects: the $N$-representability of the
functional is not explicitly treated because the numerical set up is such that the
GSPI QMC point of view implicitly assumes to work with a (numerically) exact
$3N$-dimensional wavefunction; a more rigorous formalization, however, should take
into account also the question of $N$-representability of the functional.
\section{The method at work: A wish list}
At this point, the most important technical question regards the numerical
realization of the key aspect of the formal procedure: how to implement within the
GSPI QMC a sampling procedure at given electron density $\rho({\bf r})$. This in
turn allows, {\it on-the-fly},
the determination of the correct energy functional (for the given density) within a
DFT minimization problem. This would lead to a second step, that is to construct a
numerical iterative procedure for the solution of the DFT minimization problem,
linked to the GSPI QMC calculation of the energy functional. In practice one can
envisage the procedure as a combination of a standard GSPI QMC calculation and an
energy minimization of a DFT problem. The potential technical advantage is that the
sampling at given $\rho$, within the GSPI QMC procedure, may be efficiently done by
using the pre-selection of configurations provided by the predefined target density.
The density is then updated by the DFT minimization; this implies that the resulting
GSPI QMC scheme may be highly simplified; if this is true than we would have a 
speed up for QMC calculations. In fact, within the standard QMC one must sample a
very large number of configurations
corresponding to many different densities because implicitly one must also sample
$\rho({\bf r})$ in order to reach the one of ground state. Instead, as underlined
above, having a discrete set of
predefined densities one has a fast pre-selection criterion for
the spatial configurations to be sampled.
On the other hand, from the point of view of the DFT problem, one would certainly
increase the cost of the calculations (compared to standard DFT schemes) because at
each iterative step of DFT a QMC calculation is required, but it would be sure of
having a highly accurate functional without specifying any empirical parameters or
set of parameters {\it a priori}. This can be seen as a reasonable compromise
between QMC (with its high accuracy) and DFT (with its high efficiency) and thus
represent a novel approach to go beyond KS DFT based techniques. The key point is to
make the computational code as efficient as possible so that indeed one can apply
the method in those cases where the accuracy needed is such that the standard DFT
approach cannot assure it, but standard QMC calculations would be computationally
too demanding. However, even in case the computational balance would not be
advantageous, this approach remains a tool for deriving accurate energy functionals
which can then be used in the development of exchange and correlation functionals in
a complementary fashion with respect to the standard energy functional designing
techniques. For example producing data sets for Machine-Learning-based procedures
adapted to the designing of functionals as mentioned in the chapter of von Lilienfeld.
In conclusion: if one can build a very efficient algorithm for sampling within the
GSPI QMC at constant $\rho({\bf r})$, together with an efficient algorithm for the
solution of the DFT variational problem, then the approach described so far would
represent a new path to improve the accuracy of DFT calculation at high, but most
likely reasonable, computational cost (at least for critical systems). Below we
describe some basic ideas along the direction of the development of a practical
algorithm.   

\subsection*{Specific quantities to calculate}
Despite the formalism described before may give the impression that technically the
calculations involved are massive and complicated, in reality we need the elaborated
formalism only for the conceptual justification of the iterative procedure. However the actual
calculations are practically much simpler and we do not need to explicitly calculate
$f$, as explained below.
The term $C({\bf r})=\int_{{\bf R}^{N-1}}\frac{f({\bf r}_{2},....{\bf
  r}_{N}|{\bf r})}{|{\bf r}-{\bf r}_{2}|}d{\bf r}_{2}....d{\bf r}_{N}$, can be
efficiently calculated in the standard way it is currently done in the GSPI QMC
approach, that is, the density is determined as the number of electrons visiting
some volume elements from the middle time slice, it follows that $C({\bf r})$ is
the average electron-electron potential in that volume element.\\
What is not standard is the determination of the non-local kinetic energy term:
 $I({\bf r})=\int_{{\bf R}^{N-1}}\frac{|\nabla_{{\bf r}}f({\bf r}_{2},....{\bf
  r}_{N}|{\bf r})|^{2}}{f({\bf r}_{2},....{\bf
  r}_{N}|{\bf r})}d{\bf r}_{2}....d{\bf r}_{N}$.\\
What we exactly need is:
\begin{equation}
I_{f_{GS}}({\bf r})=\int_{{\bf R}^{N-1}}\frac{|\nabla_{{\bf r}}f_{GS}({\bf
r}_{2},....{\bf
  r}_{N}|{\bf r})|^{2}}{f_{GS}({\bf r}_{2},....{\bf
  r}_{N}|{\bf r})}d{\bf r}_{2}....d{\bf r}_{N}
\end{equation}
At this point one can notice that:\\
$\left\langle K\right\rangle=\int K_{GS}({\bf r})d{\bf r}=\int\rho_{GS}({\bf
r})\left[\frac{1}{8}I_{f_{GS}}({\bf
r})+\frac{1}{8}\frac{|\nabla\rho_{GS}|^{2}}{\rho^{2}_{GS}}\right]d{\bf r}$.\\
Where $\left\langle K\right\rangle$ is the average kinetic energy of the system (which should be
naturally calculated in the QMC procedure) and $K_{GS}({\bf r})$ is the average
kinetic energy of the system as a function of the position in space.\\
The QMC procedure can calculate $K_{GS}({\bf r})$ on a grid in a straightforward
way, then $I_{f_{GS}}({\bf r})$ is automatically calculated as
$\frac{1}{\rho_{GS}}K_{GS}({\bf
r})-\frac{1}{8}\frac{|\nabla\rho_{GS}|^{2}}{\rho^{2}_{GS}}$.\\
Since $\rho_{GS}({\bf r})$, at the $i-th$ step in the iterative procedure corresponds to $\rho^{i}_{trial}({\bf r})$ and it is fixed in the QMC procedure, what is needed from QMC is only $K_{GS}({\bf
r})$.\\
It must be noticed that the kinetic functional and in particular the correlation
part, which could be easily calculated with our approach, is {\it per se} a rather
interesting subject, source of a rather active debate within the community of
orbital free DFT, as reported in the chapter of Hamilton and and in that of
Karasiev {\it et al.}\\

\subsection*{QMC Sampling at given $\rho$: Skeleton of the Algorithm}
The sampling at given $\rho({\bf r})$ in QMC is, most probably, the crucial point of
the procedure.
Following the formal derivation of the previous section, what is actually required
within the GSPI QMC procedure is the sampling of an interacting gas of electrons at
a given {\it ``forced''} density.
The sampling at given density can be done in the following way: the average one-particle density $\rho(\textbf{r})$ is evaluated during the the GSPI sampling, e.g.
by binning the
positions of the beads in a suitable 3D grid. Each MC move is accepted provided that,
besides passing the usual test of the GSPI algorithm, the new density is closer to
the target density
in a MC sense. This means that the move is always accepted if the new density
$\rho_n$ is closer to the target (trial) density than the old density $\rho_o$, it
is accepted with a certain probability if the new density departs from the target
density. This is done, for instance, by using the (square of the) Euclidean distance
between the two densities, $\mathcal{D}^2(\rho,\rho_{trial}) = \int d\textbf{r}
(\rho(\textbf{r})-\rho_{trial}(\textbf{r}))^2 \sim \sum_i (\rho(\textbf{r}_i) -
\rho_{trial}(\textbf{r}_i))^2 $, where the summation goes over the grid points. If
$\mathcal{D}^2(\rho_n,\rho_{trial}) \langle \mathcal{D}^2(\rho_o,\rho_{trial})$ the move
is always accepted, otherwise it is accepted if a random number taken from a uniform
distribution between 0 and 1 is smaller than $\exp [k
(\mathcal{D}^2(\rho_o,\rho_{trial}) - \mathcal{D}^2(\rho_n,\rho_{trial})]$ where $k$
is a suitable weight chosen such that the acceptance is neither too high nor too
low. 
This strategy is similar to the parallel tempering umbrella sampling used by Auer
and Frenkel \cite{auer}; in that case the tethered quantity was the crystal size,
here it is the one-particle density. In the same spirit of this reference, one can more
efficiently sample the new rho $\rho_n$ by accumulating over few MC regular moves,
then a complete set of moves is accepted or rejected on the basis of the test on
$\rho$. The length of the trajectory over which each new evaluation of $\rho_n$ is
performed has to be tuned for an efficient sampling.

The constrained sampling could be done as follows: in the QMC procedure the electron
density is calculated as the number of visits an electron makes to a volume element,
this means that if we sample global electron configurations with a density
constraint, we would know a priori how to generate ``reasonable'' configurations by
avoiding that a volume element is visited a number of times larger that the one
dictated by the density constraints. In practice, since the GSPI QMC approach is
mapped onto some sort of simplified linear polymer melt, the generation of melt
configurations can be ``driven'' a priori by making sure that a volume element is
not visited by a polymer more than the density constraints allows for. This avoids
the need to use the expensive machinery of a full QMC procedure for the rejection of
a ``bad'' configuration. In practice giving a set of sequential discrete densities
as constraints, part of the $3N$-dimensional problem of the QMC is reduced to a
$3$-dimensional problem of DFT.  
The overall result is that this procedure, if optimized, may restrict enormously the
configuration space to explore thus reducing the costs of the QMC procedure.\\
Next, once the energy functional at a given $\rho$ is obtained, we use it as a {\it
``energy functional guess''} for a DFT minimization. This latter implies the
solution of a standard Euler equation in three dimensions. The resulting
computational procedure will then come out by merging in a consistent loop these
different parts.  
\subsection{Helium Dimer: An Ideal Test System}
 The Helium dimer is a computationally convenient system due to the relatively
modest computational costs in QMC, compared to other systems. Moreover it is a
system used often as a prototype for understanding the physical insights of the van
der Waals interactions for bonded systems and thus there is a ``good'' literature
of reference available, from both QMC and DFT point of view (see
e.g.Refs.\cite{hel1,hel2,hel3}).
For the reasons above, applying our idea to the helium dimer would be technically
optimal, one would have reference data to compare results and may be even able to
provide novel insights into the physics of the van der Waals interactions by, for
example, identifying the contribution of the correlation term of the kinetic part of
the energy functional and compare it to the Coulomb contribution; the balance
between these terms is a question that is still open and that now can be addressed
in concrete terms.
 
\section{Conclusions}
The exploration of novel paths to the design of energy functionals for
computationally accurate many-electron approaches is a mandatory task for the next
decades. 
Novel must not necessarily mean ground-breaking exotic ideas or exact methods which
are substitutive of current ones, it implies instead constructive complementarity to
the available techniques and ideas; a decisive step forward
can only come from a multidisciplinary effort: optimal combination of physical well-founded principles, mathematical rigor, computational efficiency and chemical
accuracy. This in turn requires the capability to merge ideas and techniques
which may not look, at a first glance, compatible and thus may lead to skepticism
from each separated community. 
The idea reported in this chapter attempts to merge the (only) fundamental principle
of DFT with the idea of stochastic sampling of QMC. Indeed what we propose is the
essence of DFT, that is a $3N$-dimensional wavefunction (sampled implicitly by QMC)
as a function(al) of the three dimensional electron density.  

\end{document}